\documentclass[superscriptaddress,reprint,citeautoscript,prx]{revtex4-2} 
\usepackage{graphicx}
\usepackage[margin=0.7in]{geometry}
\usepackage{amsmath}
\usepackage{hyperref}
\usepackage{multirow}
\usepackage{newtxtext}                                        
\usepackage{booktabs}
\usepackage[dvipsnames]{xcolor}
\usepackage{hyperref}
\usepackage[normalem]{ulem}
\usepackage[textwidth=1.5cm,textsize=footnotesize]{todonotes}
\hypersetup{colorlinks, 
	linkcolor={blue!75!black!80!yellow},
	citecolor={blue!75!black!80!yellow}, 
	urlcolor={blue!75!black!80!yellow}
	}
\usepackage[cmintegrals]{newtxmath}
\usepackage{lipsum}
\usepackage{float}

\newcommand{\ECEPrinceton}{Department of Electrical and Computer Engineering, Princeton University, Princeton, New Jersey 08540, USA}
\newcommand{\PhysPrinceton}{Department of Physics, Princeton University, Princeton, New Jersey 08540, USA}
\newcommand{\ChemPrinceton}{Department of Chemistry, Princeton University, Princeton, New Jersey 08540, USA}
\newcommand{\Stanford}{Stanford Institute for Materials and Energy Sciences, SLAC National Accelerator Laboratory, Menlo Park, California 94025, USA}
\newcommand{\IAC}{Princeton Materials Institute, Princeton University, Princeton, New Jersey 08540, USA}
\newcommand{\BNL}{National Synchrotron Light Source II, Brookhaven National Laboratory, Upton, New York 11973, USA}
\AtBeginDocument{%
    \newwrite\bibnotes
    \def\bibnotesext{Notes.bib}
    \immediate\openout\bibnotes=\jobname\bibnotesext
    \immediate\write\bibnotes{@CONTROL{REVTEX41Control}}
    \immediate\write\bibnotes{@CONTROL{%
    apsrev41Control,author="08",editor="1",pages="1",title="0",year="1"}}
     \if@filesw
     \immediate\write\@auxout{\string\citation{apsrev41Control}}%
    \fi
}%

\begin{document}
\author{Faranak Bahrami}\affiliation{\ECEPrinceton}
\author{Matthew P. Bland}\affiliation{\ECEPrinceton}
\author{Nana Shumiya}\affiliation{\ECEPrinceton}
\author{Ray D. Chang}\affiliation{\ECEPrinceton}
\author{Elizabeth Hedrick}\affiliation{\ECEPrinceton}
\author{Russell A. McLellan}\affiliation{\ECEPrinceton}
\author{Kevin D. Crowley}\affiliation{\PhysPrinceton}
\author{Aveek Dutta}\affiliation{\ECEPrinceton}
\author{Logan Bishop-Van Horn}\affiliation{\Stanford}
\author{Yusuke Iguchi}\affiliation{\Stanford}
\author{Aswin Kumar Anbalagan}\affiliation{\BNL}
\author{Guangming Cheng}\affiliation{\IAC}
\author{Chen Yang}\affiliation{\ChemPrinceton}
\author{Nan Yao}\affiliation{\IAC}
\author{Andrew L. Walter}\affiliation{\BNL}
\author{Andi M. Barbour}\affiliation{\BNL}
\author{Sarang Gopalakrishnan}\affiliation{\ECEPrinceton}
\author{Robert J. Cava}\affiliation{\ChemPrinceton}
\author{Andrew A. Houck}\affiliation{\ECEPrinceton}
\author{Nathalie P. de Leon}\affiliation{\ECEPrinceton}

\title{Vortex Motion Induced Losses in Tantalum Resonators}


\begin{abstract}
Tantalum (Ta) based superconducting circuits have been demonstrated to enable record qubit coherence times (T$_{2}$) and quality factors ($Q$), motivating a careful study of the microscopic origin of the remaining losses that limit their performance. We have recently shown that the losses in Ta-based resonators are dominated by two-level systems (TLSs) at low microwave powers and millikelvin temperatures. We also observe that some devices exhibit loss that is exponentially activated at a lower temperature inconsistent with the superconducting critical temperature (T$_{c}$) of the constituent film. Specifically, dc resistivity measurements show a T$_{c}$ of over 4 K, while microwave measurements of resonators fabricated from these films show losses that increase exponentially with temperature with an activation energy as low as 0.3 K. Here, we present a comparative study of the structural and thermodynamic properties of Ta-based resonators and identify vortex motion-induced loss as the source of thermally activated microwave loss. Through careful magnetoresistance and x-ray diffraction measurements, we observe that the increased loss occurs for films that are in the clean limit, where the superconducting coherence length ($\xi$) is shorter than the mean free path ($l$). Vortex motion-induced losses are suppressed for films in the dirty limit, which show evidence of structural defects that can pin vortices. We verify this hypothesis by explicitly pinning vortices via patterning and find that we can suppress the loss by microfabrication.

\end{abstract}

\maketitle

\section{Introduction}

Superconducting qubits are one of the most promising candidates for large scale quantum processors, and have been deployed in demonstrations of quantum error correction~\cite {krinner2022realizing, Gong2021Experimental,sivak2023realtime,ofek2016extending}, quantum simulation~\cite{barends2015digital, kandala2017hardware, marcos2013superconducting}, and quantum many-body physics~\cite{zhang2022synthesizing, mi2022time, kollar2019hyperbolic, andersen2024thermalization}. Building large scale processors will require improving the underlying qubit performance by identifying and mitigating losses in current state-of-the-art superconducting circuits. Recently, utilizing tantalum (Ta) as the superconducting material for capacitors improved both lifetimes (T$_{1}$) and coherence times (T$_{2}$) of superconducting qubits~\cite{place2021new, ganjam2024surpassing, wang2022towards, sivak2023realtime, gao2024establishing}. Ta is chemically robust, allowing for post-fabrication cleaning such as acid treatment to remove hydrocarbon contamination at the surface. Furthermore, Ta has a stoichiometric native surface oxide layer that is self-limiting and  can be thinned with buffered oxide etch (BOE)~\cite{crowley2023disentagling, mclellan2023chemical}, further decreasing losses associated with amorphous surface oxides. The remarkable improvements associated with Ta motivate a careful investigation of the different microwave loss channels in these circuits, including two-level systems (TLSs)~\cite{anderson1972anomalous, muller2019towards}, quasiparticles~\cite{serniak2018hot, Connolly2024coexistence}, Andreev bound states~\cite{tinkham2004introduction, sauls2018andreev}, and magnetic flux vortices~\cite{Nsanzineza2014trapping, song2009microwave, chiaro2016dielectric, stan2004critial, catelani2022ac}. 

We recently studied the power and temperature dependence of loss in Ta superconducting resonators to show that the dominant loss mechanism at low temperature and low microwave power is saturable absorption of TLSs at the surface and in the bulk substrate~\cite{crowley2023disentagling}. However, in some devices we observed an additional loss channel that is exponentially activated with temperature, with an activation energy that is inconsistent with the T$_{c}$ of the superconducting film. Originally it was hypothesized~\cite{crowley2023disentagling} that a mixture of the tetragonal $\beta$-Ta (T$_{c}$ $\sim$ 0.7 K) and the body-centered cubic (BCC) $\alpha$-Ta phase (T$_{c}$ $\sim$ 4.3 K)~\cite{schwartz1972temperature, mazin2022superconducting, schwartz1977impurity} was the source of this anomalous temperature-dependent loss. However, x-ray diffraction (XRD), transmission electron microscopy (TEM), and scanning SQUID measurements reveal no evidence of inhomogeneity in these films at the nanometer and micrometer scales, respectively, ruling out the presence of the tetragonal phase in these resonators (Appendix~\ref{TEM}, \ref{SQUID}). 

\begin{figure*}
   \centering
  \includegraphics[width=\textwidth]{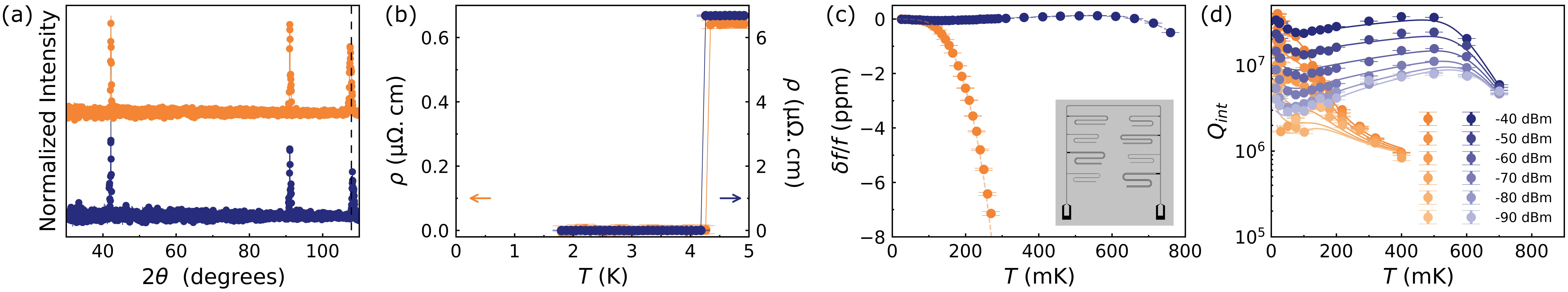}
  \caption{\textbf{Nominally Identical $\alpha$-Ta Films, Drastically Different Microwave Losses:} Comparative measurements of type A (orange) and type B (purple) Ta thin films. (a) X-ray diffractometry of representative type A and type B films used to fabricate Ta resonators. The peaks at 41.6$^{\circ}$ and 90.8$^{\circ}$ represent the sapphire substrate peaks and the peak at 107.5$^{\circ}$, represented by the dashed line, corresponds to the $\alpha$-Ta (222) peak position. A small shift is observed between the $\alpha$-Ta (222) peak positions. The patterns are vertically offset for clarity.
  (b) DC resistivity measurements as a function of temperature show a sharp drop at the superconducting transition temperature (T$_{c}$), which is 4.3~K for the type A film and 4.2~K for the type B film, confirming the BCC $\alpha$-Ta phase. The residual resistivity, however, is one order of magnitude smaller in type A (left axis) compared to type B (right axis). 
  (c) Measured fractional frequency shift as a function of temperature for resonators fabricated from type A and type B films similar to those used in (a) and (b). The dashed lines are fits to the data. The $\delta f/f$ for the type A film drops around 150~mK and suggests an activation energy as low as 0.3~K, while the type B film remains relatively constant out to 700~mK, and the extracted T$_{c}$ is above 4~K. Inset is an image of a coplanar waveguide (CPW) resonator device used in this measurement. 
  (d) Measured internal quality factor (Q$_{int}$) as a function of temperature with varying applied microwave power for resonators fabricated from type A and B films similar to those used in (a) and (b). The solid lines are fits to the data. The extracted activation temperatures for exponential, power independent decay are 0.3~K and 4~K, respectively.}
   \label{fig:Puzzle}
\end{figure*}

Here, we conduct a detailed investigation of the structural and transport properties of Ta-based resonators to elucidate the microscopic origins of microwave loss. Although both XRD and dc resistivity confirm the $\alpha$-Ta phase in all devices (Fig.~\ref{fig:Puzzle}(a, b)), we observe two classes of films with key differences that correlate with anomalous microwave loss as measured by the temperature-dependent frequency shift (Fig.~\ref{fig:Puzzle}(c)) and quality factor (Fig.~\ref{fig:Puzzle}(d)). Specifically, the two classes of films show residual resistivity ratios (RRRs) that differ by approximately a factor of five on average, and lattice constants that differ by 0.013~\AA, suggesting that they exhibit different defect densities. We determine that the Ta thin films used to fabricate these resonators can be categorized into two main types: films with fewer defects that are in the superconducting clean limit (type A), and films with more defects that are in the dirty limit (type B). We hypothesize that the defects in the type B films pin vortices, increasing the activation energy for vortex motion and thereby reducing microwave loss. We verify this hypothesis by explicitly pinning vortices in a type A film via microfabrication.

\section{Clean and Dirty Limits of Superconductivity}
\begin{figure}
   \centering
  \includegraphics[width=0.4\textwidth]{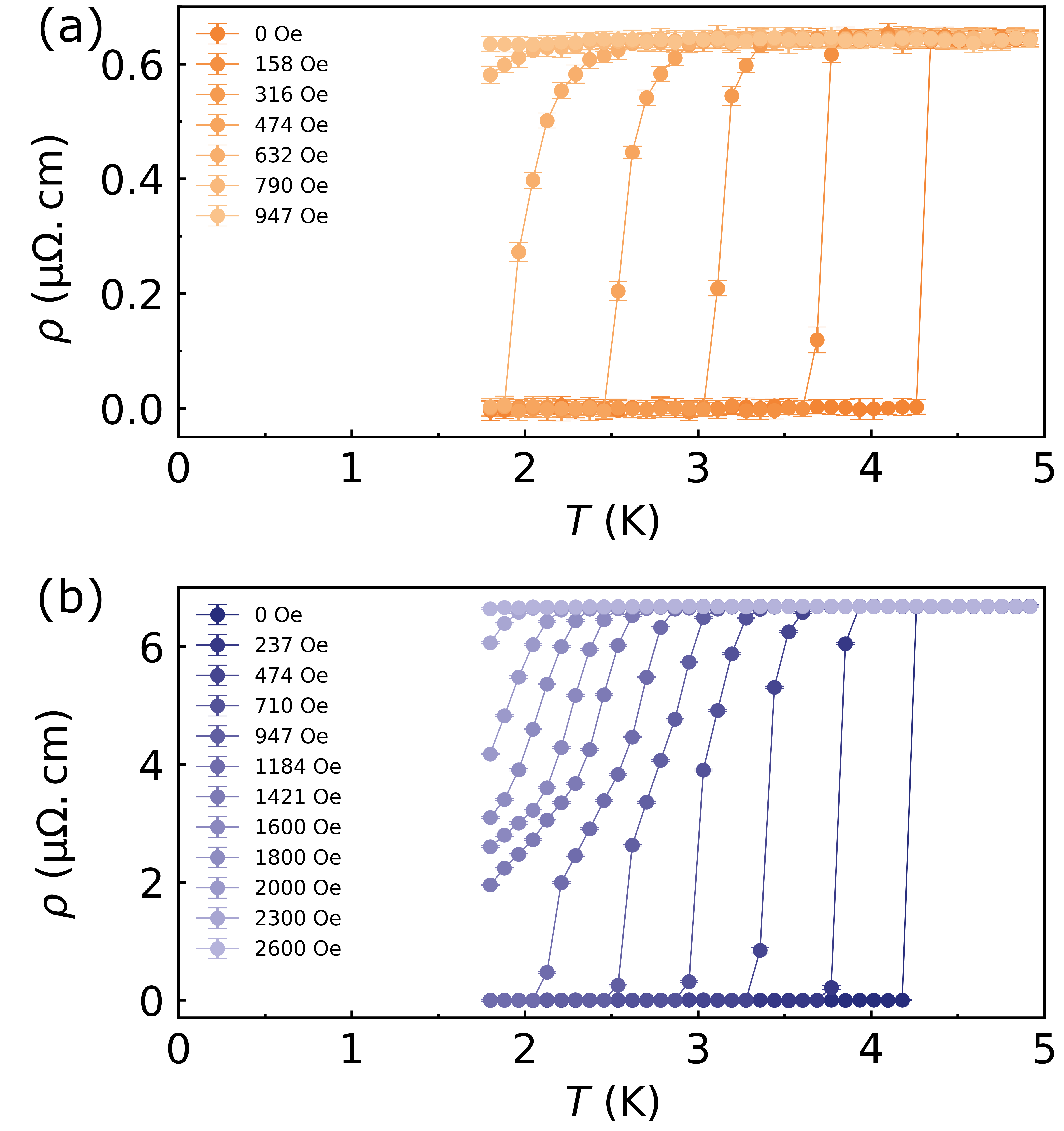}
  \caption{\textbf{Magnetoresistivity:} DC resistivity as a function of temperature below 5~K with varying out-of-plane applied magnetic field for type A (orange) and type B (purple) films. The strength of the applied magnetic field needed to suppress superconductivity is significantly larger for type B films.
  }
   \label{fig:2}
\end{figure}

We observe that the Ta thin films are type-II superconductors. Type-II superconductors host a mixed state, in which magnetic fields are able to penetrate the superconductor in localized regions known as vortices, where the supercurrent for each vortex gives rise to a single quantum of magnetic flux: $\Phi_{0} \equiv \frac{h}{2e}$. The motion of vortices can lead to energy dissipation at microwave frequencies~\cite{mcrae2020materials, song2009microwave, chiaro2016dielectric, stan2004critial}, and vortex motion is suppressed if the vortices are pinned by defects in the film~\cite{Nsanzineza2014trapping, song2009microwave, chiaro2016dielectric}. The mixed state is bounded between two critical fields: the lower-limit critical field, H$_{c1}$, and the upper-limit critical field, H$_{c2}$~\cite{abrikosov1957magnetic, Bardeen1965theory, matsuda2002free}.
Both H$_{c1}$ and H$_{c2}$ carry important information about the characteristic properties of vortices in the mixed state. H$_{c1}$ is proportional to the inverse square of the London magnetic penetration depth, H$_{c1} \propto \lambda^{-2}$. Similarly, H$_{c2}$ is proportional to the inverse square of the coherence length, H$_{c2} \propto \xi^{-2}$, which characterizes the size of the normal metal region within the core of each vortex~\cite{rainer1996current, womack2021extreme}.

\begin{table*}[t]
\centering
\caption{\label{table:transport_comparison} Summary of activation temperature (T$_{act}$), critical temperature (T$_{c}$), upper limit critical field (H$_{c2}$), coherence length ($\xi$), mean free path ($l$), residual resistivity ratio (RRR), normal resistivity ($\rho_{n}$) at 5~K, and viscous drag coefficient ($\eta$) for each measured device.}
\begin{ruledtabular}
\begin{tabular}{ccccccccccc}
   & Device & T$_{act}$~(K) & T$_{c}$~(K) & H$_{c2}$(0)~(T) & $\xi$~(nm)& $l$~(nm) &   Limit & RRR & $\rho_{n}$ (5~K)~ ($\mu$$\Omega \cdot$cm)& $\eta$~$\times$ $10^{-8}$~ (kg/m$\cdot$s)  \\
       \hline
       \parbox[t]{1mm}{\multirow{5}{*}{\rotatebox[origin =c]{90}{Type A}}}
       &    D$_{1}$	&   0.54 $\pm$ 0.04    &   4.33 $\pm$ 0.1   &    0.108 $\pm$ 0.002   &   55.3 $\pm$ 0.5    &  109.3 $\pm$ 0.8    &   Clean  & 39.5 $\pm$ 0.2  &   0.77 $\pm$ 0.01   &   2.89 $\pm$ 0.06    \\
       &    D$_{2}$	&   0.37 $\pm$ 0.03   &   4.39 $\pm$ 0.1    &    0.132 $\pm$ 0.002   &   49.9 $\pm$ 0.4    &   142.3 $\pm$ 0.2     &   Clean  & 65.1 $\pm$ 0.6  &   0.55 $\pm$ 0.01   &   4.99 $\pm$ 0.08   \\
       &    D$_{3}$	&   0.29 $\pm$ 0.01   &   4.33 $\pm$ 0.1    &    0.165 $\pm$ 0.005   &   44.6 $\pm$ 0.6    &   81.5 $\pm$ 0.7   &   Clean  & 38.5 $\pm$ 0.3  &   0.87 $\pm$ 0.01   &   3.91 $\pm$ 0.12   \\
       &    D$_{4}$	&   0.33 $\pm$ 0.01    &   4.28$\pm$ 0.1    &    0.166 $\pm$ 0.003   &   44.5 $\pm$ 0.4    &   125.2 $\pm$ 1   &   Clean  & 62.2 $\pm$ 0.7  &   0.65 $\pm$ 0.01   &   5.33 $\pm$ 0.11   \\
       &    D$_{5}$	&   0.57 $\pm$ 0.01    &   4.30 $\pm$ 0.1    &    0.099 $\pm$ 0.001   &   57.4 $\pm$ 0.2   &  112.2 $\pm$ 2    &   Clean  & 40.8 $\pm$ 0.8  &   0.65 $\pm$ 0.02   &   3.23 $\pm$ 0.07   \\
       \hline\hline
       \parbox[t]{1mm}{\multirow{4}{*}{\rotatebox[origin =c]{90}{Type B}}}
       &    D$_{6}$	&   1.69 $\pm$ 0.36 &   4.22 $\pm$ 0.1    &    0.199 $\pm$ 0.001   &   40.6 $\pm$ 0.1    &   20.4 $\pm$ 0.1   &   Dirty  & 11.5 $\pm$ 0.05  &   2.61 $\pm$ 0.02    &   1.58 $\pm$ 0.01    \\
       &    D$_{7}$	&   4.16 $\pm$ 0.04 &   4.22 $\pm$ 0.1    &    0.245 $\pm$ 0.004   &   36.6 $\pm$ 0.3   &   12.9 $\pm$ 0.04   &   Dirty  & 10.8 $\pm$ 0.03 &   6.68 $\pm$ 0.02    &   0.755 $\pm$ 0.01  \\
       &    D$_{8}$	&   4.17 $\pm$ 0.28 &   4.33 $\pm$ 0.1    &    0.407 $\pm$ 0.001  &   28.4 $\pm$ 0.1    &   13.3 $\pm$ 0.02    &   Dirty  & 6.78 $\pm$ 0.01  &   5.31 $\pm$ 0.01    &   1.59 $\pm$ 0.01
   \end{tabular}
   \end{ruledtabular}
\end{table*}

We measure the resistivity as a function of temperature and applied magnetic field for type A and B films to extract H$_{c2}$. For a representative type A film at zero applied field the superconducting transition is observed at 4.3~K (Fig.~\ref{fig:2}(a)), as expected for BCC $\alpha$-Ta. With increasing applied field the superconducting phase is suppressed, until at 947~Oe there is no sign of a transition in the resistivity down to 1.8~K. Similarly, a representative type B film exhibits a transition temperature of 4.2~K at 0~Oe; however, achieving the same suppression in superconductivity requires a significantly larger applied field of 2600~Oe (Fig.~\ref{fig:2}(b)). This behavior indicates that H$_{c2}$ is smaller for the type A films compared to the type B films. To determine the upper-limit critical field at T = 0~K, H$_{c2}(0)$, we extract the transition temperature T$_{c}$ at each applied field and plot the magnetic field vs. T$_{c}$ phase diagram for each film (Fig.~\ref{fig:Hc2vsT}). The results are modeled using Ginzburg-Landau theory~\cite{cyrot1973ginzburg, kushwaha2024superconductivity},
\begin{equation}
\label{GL equ}
    H_{c2} (T) = H_{c2} (0) \left[\frac{1 - t^{2}}{1 + t^{2}} \right], \qquad  t = \frac{T}{T_{c}}
\end{equation}
where $t$ is the reduced temperature (Appendix~\ref{Hc2}). The coherence length $\xi$ for each film can be obtained according to H$_{c2} (0) = \frac{\Phi_{0}}{2\pi \xi^{2}}$~\cite{barker2018superconducting}. Type B films exhibit larger H$_{c2}(0)$ values than for type A films. Accordingly, the coherence length is smaller in type B films compared to type A films (Table~\ref{table:transport_comparison}).

We compare the coherence length to the mean free path, $l$, for each film. The mean free path is extracted from Hall measurements of the Ta thin films (Appendix~\ref{Hall}). Type A films show mean free paths with an average of $\sim$114~nm, while the type B films show shorter mean free paths with an average of $\sim$15~nm  (Table~\ref{table:transport_comparison}). Comparing the coherence length and mean free path for each device reveals a consistent trend: type A films have $\xi < l$, in the clean superconducting limit, while type B films have $\xi > l$ and are in the dirty limit, which suggests different densities of defects in the two categories of films.

One indicator of the defect density is the RRR, which is the ratio of the resistivity at room temperature to the resistivity just above the superconducting transition (i.e., the normal state resistivity, $\rho_{n}$ (5~K)). A higher RRR generally indicates a lower density of defects. The normal state resistivity itself also provides an indication of the level of defects, with a lower $\rho_{n}$(5~K) corresponding to a lower density of defects. Type A films have an average RRR value of $\sim$49 with $\rho_{n} (5~$K$)<1$~$\mu\Omega\cdot$cm, while type B films have an average RRR value of $\sim$9.7 with $\rho_{n} (5~$K$) > 2$~$\mu\Omega\cdot$cm (Table~\ref{table:transport_comparison}). The higher RRR and lower $\rho_{n}$ (5~K) values for the type A films relative to type B films are consistent with the classification into clean and dirty limits provided by the coherence lengths and mean free paths.

Vortices are typically pinned at the sites of defects; their presence alone does not necessarily contribute to microwave loss. The motion of vortices, however, can give rise to dissipation at microwave frequencies~\cite{mcrae2020materials, Nsanzineza2014trapping, song2009microwave, chiaro2016dielectric, stan2004critial}. A simplified model~\cite{tinkham2004introduction} of vortex motion estimates the viscous drag coefficient as:
\begin{equation}
    \eta = \frac{\Phi^{2}_{0}}{2\pi \xi^{2} \rho_{n}}.
\end{equation}
Calculating $\eta$ for our films, we find that the viscous drag coefficient for type A films is 3 times larger (on average) compared to the type B films (Table~\ref{table:transport_comparison}).

\begin{figure}
   \centering
  \includegraphics[width=0.47\textwidth]{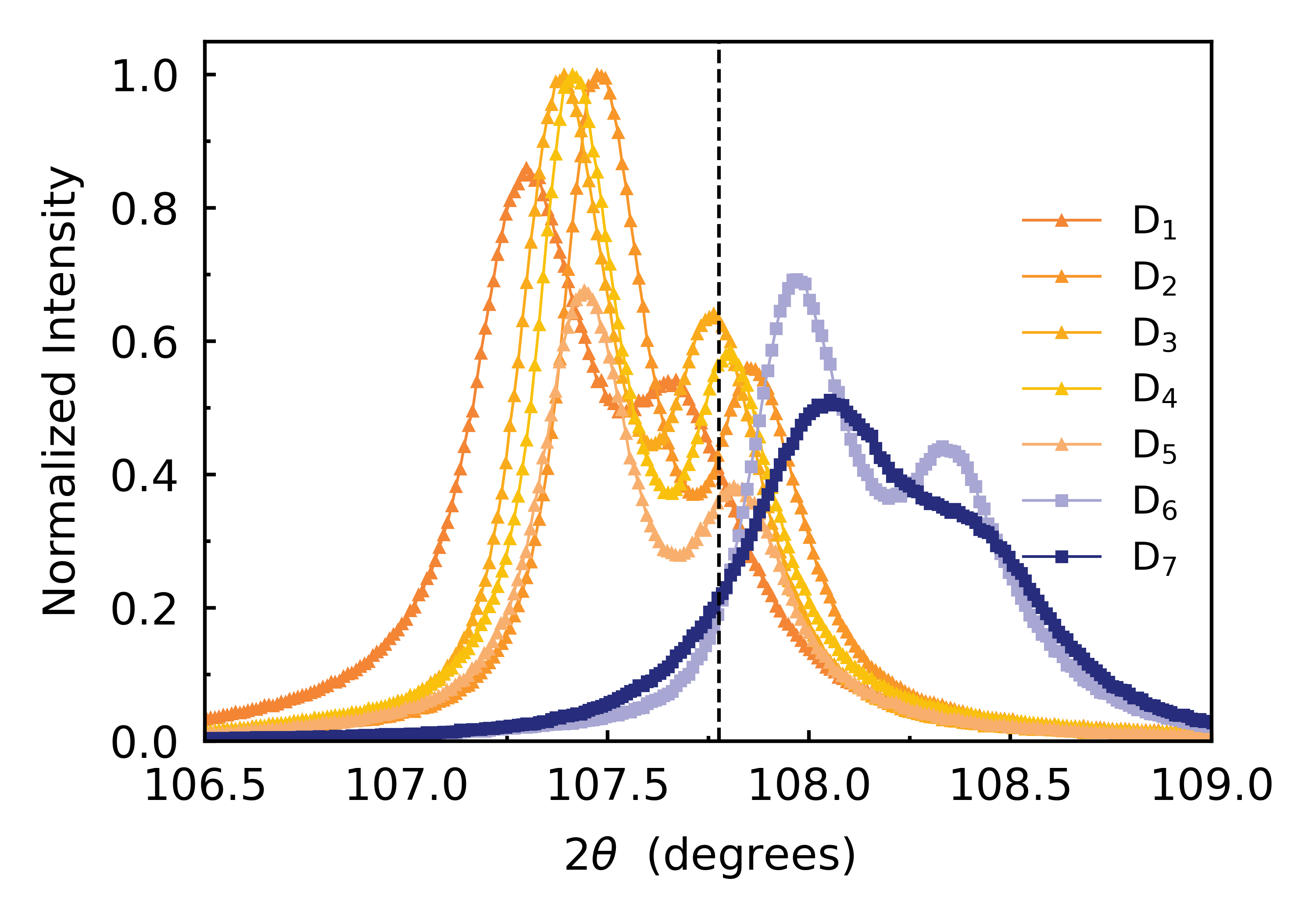}
  \caption{\textbf{Detailed structural characterization:} X-ray diffractometry of type A (orange) and type B (purple) Ta thin films around the $\alpha$-Ta (222) peak position. The dashed line represents the expected $\alpha$-Ta (222) peak position~\cite{edwards1951high}. There is a $\sim$0.7$^{\circ}$ shift to higher angles in type B films compared to type A films.}
   \label{fig:3}
\end{figure}

We now turn to a comparative structural characterization of the two types of films. The XRD patterns for type B films show a small shift to higher angles of the $\alpha$-Ta (222) peak near 107.5$^{\circ}$ compared to type A films (Fig.~\ref{fig:3}). These differences correlate primarily with the Ta deposition rate, within the relevant temperature range for $\alpha$-Ta growth (Appendix~\ref{growthcondition}). This shift is about 0.7$^{\circ}$ and suggests a small reduction in the lattice constant ($a$) of type B films compared to type A films (Table~\ref{table:lattice mismatch}). The average difference in lattice constant between type A and type B films is about 0.013~\AA, and the lattice constants lie on opposite sides of the reported value for bulk Ta from Ref.~\cite{edwards1951high} ($a_{\text{ref}}$ = 3.303~\AA, dashed line in Fig.~\ref{fig:3}). We compare these lattice parameters to the underlying substrate to characterize the heteroepitaxial lattice mismatch. The single-crystal Ta films grow epitaxially on c-axis sapphire with a $<$111$>$ growth direction. TEM measurements show that Ta along its ($1\bar{1}2$) plane and sapphire along its ($1\bar{1}00$) plane overlap at the metal-substrate interface (Appendix~\ref{TEM}). The relevant interatomic spacings for the lattice mismatch are thus d$_{1\bar{1}2}$ for Ta and d$_{1\bar{1}00}$ for sapphire~\cite{dehm1995growth}. Comparing type A and type B films (Table~\ref{table:lattice mismatch}), we see that the lattice mismatch is slightly larger in type B films than in type A films, which likely leads to a higher structural defect density. We also observe a lower degree of Ta-Ta bond ordering~\cite{yang2012local} in type B films as measured by X-ray absorption spectroscopy (XAS) at the Ta $L_{3}$ edge (Appendix~\ref{XAS}). We note that one of the type B films exhibits a different crystal structure, with growth oriented along the $<$110$>$ direction. For such films the resulting columnar grain structure likely serves as a source of defects that can also pin vortices (Appendix~\ref{110 orientaion}).

\begin{table}[h]
\centering
\caption{\label{table:lattice mismatch} Out-of-plane film orientation, lattice constant, and lattice mismatch between Ta and the sapphire substrate for each device.}
\begin{ruledtabular}
\begin{tabular}{ccccc}
   & Device & $<$hkl$>$ & Lattice Constant (\AA) & Lattice Mismatch ($\%$)   \\
       \hline
       \parbox[t]{1mm}{\multirow{5}{*}{\rotatebox[origin =c]{90}{Type A}}}
       &    D$_{1}$	&    $<$111$>$   &  3.3126(2)   &   1.449\\
       &    D$_{2}$	&    $<$111$>$   &  3.3093(2)   &   1.564\\
       &    D$_{3}$	&    $<$111$>$   &  3.3113(2)   &   1.504\\
       &    D$_{4}$	&    $<$111$>$   &  3.3107(2)   &   1.522\\
       &    D$_{5}$	&    $<$111$>$   &  3.3101(2)   &   1.539\\
       \hline\hline
       \parbox[t]{1mm}{\multirow{4}{*}{\rotatebox[origin =c]{90}{Type B}}}
       &    D$_{6}$	&    $<$111$>$   &  3.2989(2)   &   1.877\\
       &    D$_{7}$	&    $<$111$>$   &  3.2974(2)   &   1.924\\
       &    D$_{8}$	&    $<$110$>$   &  3.2959(8)   &  --
   \end{tabular}
   \end{ruledtabular}
\end{table}

\section{Vortex-Defect Interactions}
\begin{figure}
   \centering
  \includegraphics[width=0.4\textwidth]{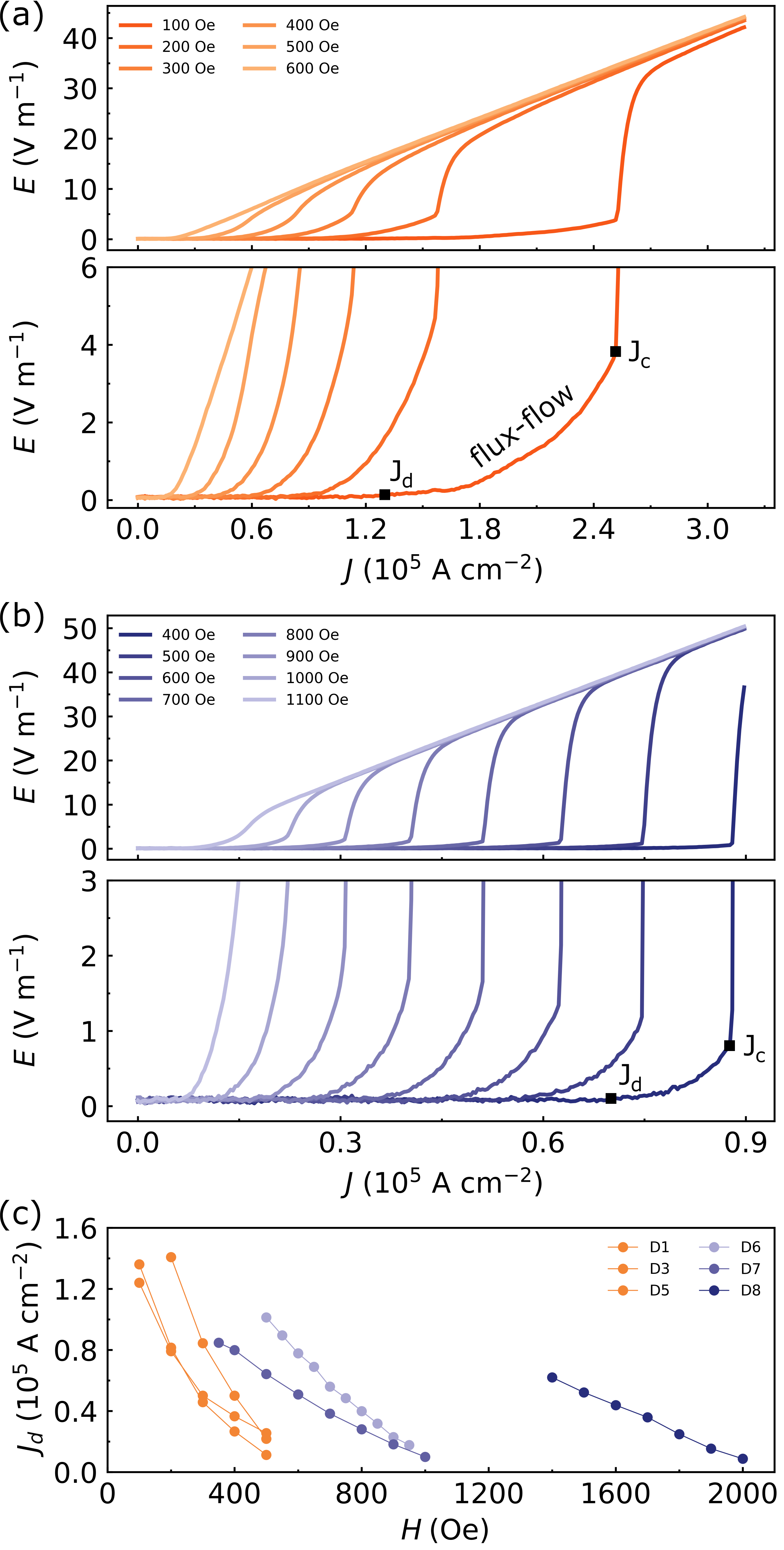}
  \caption{\textbf{E-J Characteristic:} Electric field (E) as a function of current density (J) measured at $\sim$ 50$\%$ of the T$_{c}$, 2.1~K. (a) The flux-flow regime where vortex motion occurs for a type A film. The bottom panel is zoomed-in on the flux-flow region. J$_{d}$ indicates the depinning current, and J$_{c}$ indicates the superconducting critical current. (b) E-J curves for a type B film. Flux-flow occurs at higher magnetic fields for the type B films.
  (c) J$_{d}$ as a function of applied field extracted from the E-J curves. The depinning current for a similar applied field is smaller for type A films compared to type B films due to the presence of fewer defects in type A films.}
   \label{fig:I-V}
\end{figure}
To confirm the presence of vortices and measure their dynamics in the two types of films, we measured the electric field ($E$) as a function of current density (J) under different applied magnetic fields at $\sim$50$\%$ of the transition temperature of the films, 2.1~K (Fig.~\ref{fig:I-V}(a, b)). The E-J characteristic curves are presented instead of voltage (V) vs. current (I) to account for the effects of film geometry. The E-J curves show three distinct regions: at low J values there is no resistive loss and the electric field (voltage) remains zero. For J values larger than the depinning current J$_{d}$, the Lorentz force induces vortex motion, which gives rise to small but non-zero resistance. When J is larger than the critical current (J$_{c}$), the superconducting phase is suppressed and the E-J curve exhibits a linear dependence characteristic of normal metal behavior.   

The non-linear E-J behavior between J$_{d}$ and J$_{c}$ is a hallmark of vortex motion (``flux-flow")~\cite{zhang2020quasiparticle}. We observe this flux-flow region in both types of films (Fig.~\ref{fig:I-V}(a, b)), and we observe that a smaller applied magnetic field is required to suppress the depinning force in type A films compared to type B films. We quantify this difference by plotting J$_{d}$ as a function of the applied field (Fig.~\ref{fig:I-V}(c)), for the two types of films. These results suggest that a smaller Lorentz force is required to activate the vortex motion in type A films compared to type B films, consistent with the lower activation energy of type A films.

\begin{figure}
   \centering
  \includegraphics[width=0.35\textwidth]{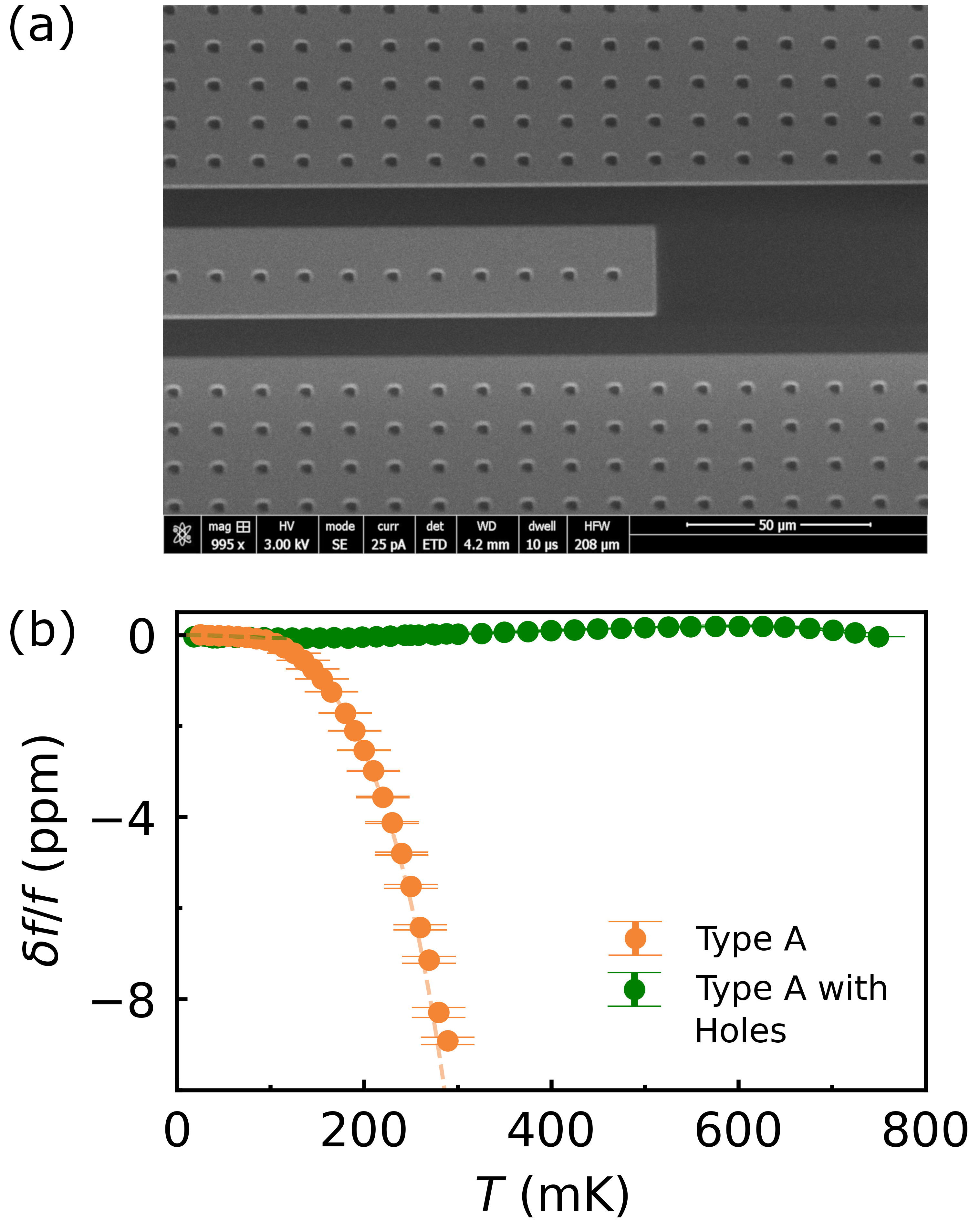}
  \caption{\textbf{Vortex Pinning:} (a) SEM image of device fabricated out of a type A film with holes introduced to suppress vortex motion. (b) Frequency shift $\delta f/f$ with temperature for a resonator fabricated from a type A film without holes (orange) and with holes (green). The onset of exponential loss is around $\sim$0.3~K for the unstructured type A film, while the micro-structured type A film has an extracted T$_{c}$ of 4.2~K, consistent with the critical temperature for the BCC $\alpha$-Ta phase.}
   \label{fig:4}
\end{figure}
Our transport measurements suggest that the additional microwave loss in type A films is due to vortex motion because of an absence of vortex pinning sites related to structural defects. To confirm this hypothesis, we fabricate coplanar waveguide (CPW) quarter-wave resonators (described in Ref.~\cite{crowley2023disentagling}) on a type A Ta film with additional artificial defects in the form of micro-structured holes (Fig.~\ref{fig:4}(a)), described in (Appendix~\ref{Methods})~\cite{Nsanzineza2014trapping, song2009microwave, chiaro2016dielectric, stan2004critial}. The introduction of holes into the film dramatically changes the temperature dependent microwave behavior (Fig.~\ref{fig:4}(b)). The resonance frequency shift exhibits an activation temperature of 0.3 K without patterning and an activation temperature of 4.2 K with patterning. The latter activation energy is consistent with the superconducting critical temperature T$_{c}$, indicating that the holes induced via microfabrication have fully suppressed the anomalous vortex motion-induced loss, thus confirming our hypothesis. To measure the impact of the patterning on surface losses, we extract the inverse linear absorption due to TLSs, Q$_{TLS, 0}$, from $\delta f/f$ measurements~\cite{crowley2023disentagling} for the micro-structured device (Appendix~\ref{loss}). We find that the presence of the holes on the ground plane and resonator centerpins does not significantly affect the TLS loss in the device.

\section{Conclusion}
We have demonstrated that fabricating higher-quality, lower-defect density Ta films on sapphire can introduce an additional loss channel associated with the motion of vortices because of the elimination of vortex pinning sites. Using structural characterization and dc transport measurements, we can connect anomalous microwave losses to microscopic film properties, and we have identified subtle structural differences that dramatically affect vortex pinning. We have also shown that the vortex motion-induced loss can be easily circumvented by explicitly introducing vortex pinning sites via microfabrication. Ongoing efforts include identifying the precise threshold densities of the artificially introduced defects necessary for mitigating vortex-induced losses. More broadly, this study demonstrates that correlating microwave characterization with detailed material characterization enables the identification of microscopic sources of loss, providing strategies for optimizing superconducting device behavior. 

\section*{Acknowledgements}
We thank Kathryn A. Moler, Leonid Glazman and Michel Devoret for fruitful discussions. This work was primarily supported by the U.S. Department of Energy, Office of Science, National Quantum Information Science Research Centers, Co-design Center for Quantum Advantage (C$^{2}$QA) under Contract No. DESC0012704. The authors acknowledge the use of Princeton’s Imaging and Analysis Center (IAC), which is partially supported by the Princeton Center for Complex Materials (PCCM), a National Science Foundation (NSF) Materials Research Science and Engineering Center (MRSEC; DMR-2011750), as well as the Princeton Micro/Nano Fabrication Laboratory. Use of the NSLS-II (NIST beamline 6-BM) was supported by the DOE Office of Science User Facility operated for the DOE Office of Science by Brookhaven National Laboratory under contract  DE-SC0012704. The scanning SQUID measurements were supported by the DOE “Quantum Sensing and Quantum Materials” Energy Frontier Research Center under Grant No. DE-SC0021238.

Princeton University Professor Andrew Houck is also a consultant for Quantum Circuits Incorporated (QCI). Due to his income from QCI, Princeton University has a management plan in place to mitigate a potential conflict of interest that could affect the design, conduct and reporting of this research.


\appendix
\section{Methods}
\label{Methods}
X-ray diffraction experiments were performed using a Bruker D8 ECO X-Ray Diffractometer. The instrument is equipped with a copper x-ray source (Cu K$_{\alpha}$). Magnetoresistivity and E-J characteristic measurements were measured using a Quantum Design PPMS Dynacool instrument. 

Resonator fabrication is performed with the following steps. Hexamethyldisilazine (HMDS) vapor is deposited at 148 °C in a YES oven, followed by a photoresist layer (AZ1518), followed by a bake at 95 $^{\circ}$C for 60 seconds. The photolithography process is performed with a Heidelberg DL66+ laser writter using the following parameters: 10 mm write head, 10$\%$ focus, 25$\%$ filter, 70$\%$ intensity (dose), followed by a post-exposure bake at 110 $^{\circ}$C for 2 minutes. The resist is developed using AZ300MIF solution for 90 seconds, rinsed with water for 30 seconds, and then blown dry with nitrogen gas. The pattern is transferred to the metal film by reactive-ion etching (RIE) with chlorine chemistry and the following parameters: chamber pressure of 5.4 mTorr, a radio frequency (RF) inductively coupled plasma (ICP) source of 500~W, a RF bias source of 50~W, with chlorine flow of 5 sccm and argon flow of 5 sccm. After the etch process, the photoresist is removed in a Remover PG bath at 80 $^{\circ}$C for an hour followed by acetone and IPA sonication each for 5 minutes, respectively.

The CPW resonator shown in Fig.~\ref{fig:4}(a) consists of micro-structured holes with a 3~$\mathrm{\mu}$m side length and 12~$\mathrm{\mu}$m center-to-center spacing that were patterned and then etched onto the resonator chip design such that the feedline and all resonators were fully surrounded by the holes. A buffer of 10~$\mathrm{\mu}$m was set around the resonator and feedline features to mitigate additional surface participation from the square holes. Additionally, the 2 and 4~$\mathrm{\mu}$m CPW resonators and feedline do not have vortex pinning holes present within their centerpin.

We used a commercial microwave package (QDevil QCage.24) with an associated Au-plated printed circuit board (PCB) similar to the packaging described in Ref.~\cite{crowley2023disentagling} and mounted to the dilution refrigerator as described in Ref.~\cite{crowley2023disentagling}. 

\begin{figure}
   \centering
  \includegraphics[width=0.4\textwidth]{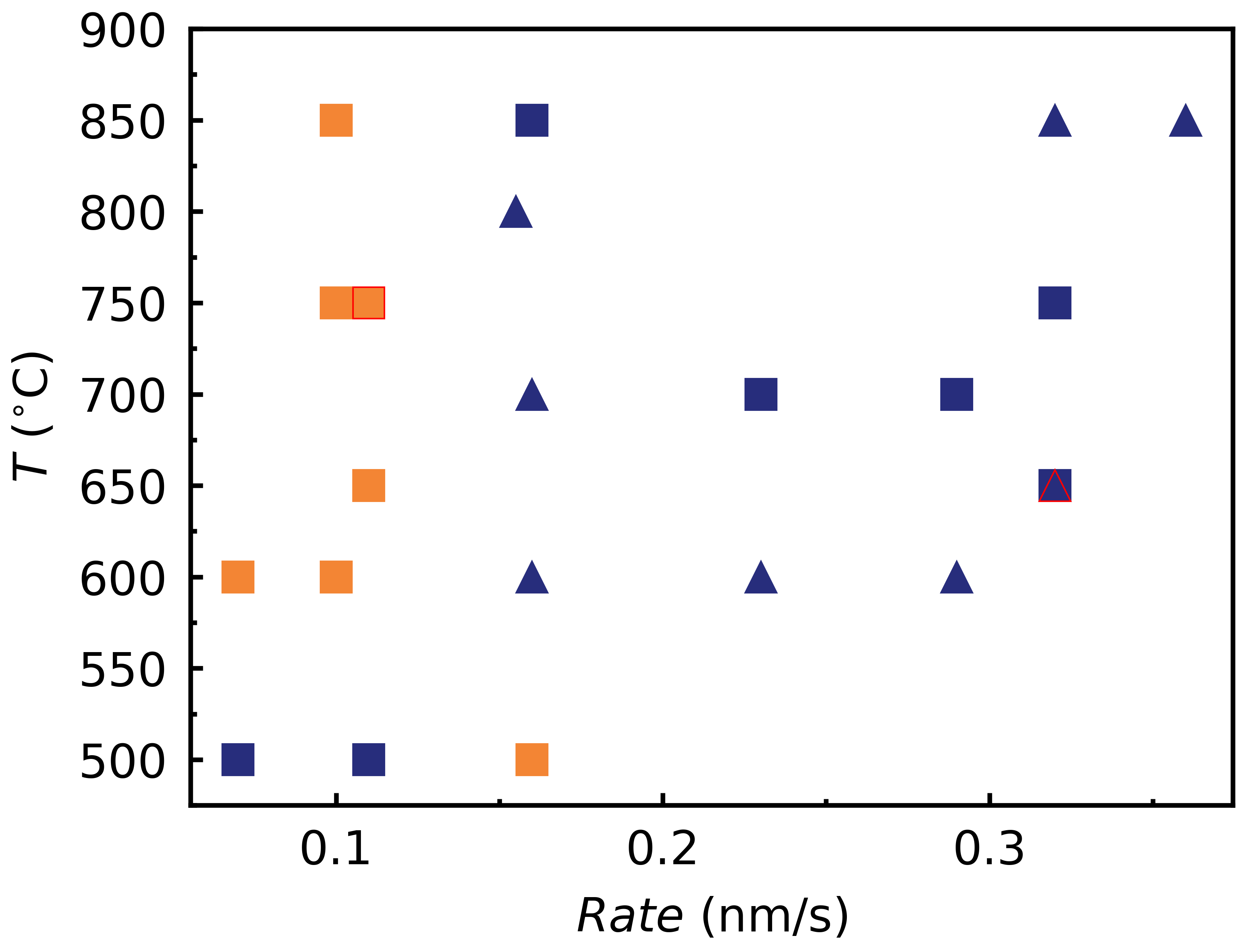}
  \caption{Ta film phase diagram for deposition temperatures between 500 $^{\circ}$C to 850 $^{\circ}$C and different deposition rates. The square and triangle symbols refers to $<$111$>$ and $<$110$>$ growth directions, respectively. The orange color refers to clean limit films and the purple color refers to the dirty limit films.}
   \label{fig:PD}
\end{figure}
\section{Growth Conditions for Clean vs. Dirty Limit Films}
\label{growthcondition}
The growth substrates are c-axis HEMEX sapphire from Crystal Systems. The wafer undergoes a cleaning process with piranha (2:1 ratio sulfuric acid: hydrogen peroxide, total 30 mL) for 20 minutes, followed by cleaning in 3 separate DI water baths followed by IPA, and is then blown dry with nitrogen gas. The wafer is then transferred into a ultra-high vacuum (UHV) dc magnetron sputtering (AJA Orion 8) chamber. The base pressure of the main chamber before starting the sputtering process is between 2$\times10^{-9}$ and 2$\times10^{-8}$ Torr.

We varied the deposition temperatures as a function of deposition rate to construct a phase diagram for preparing Ta films in the clean vs. dirty superconducting limits. The deposition temperature range varies from 500 $^{\circ}$C to 850 $^{\circ}$C for different rates (Fig.~\ref{fig:PD}). Small diced (1''$\times$ 1'') and piranha cleaned HEMEX sapphire substrates are used for this deposition phase diagram.

We precondition the chamber by depositing Ta on the chamber walls and sample holder (chuck) at room temperature, then we wait until the chamber reaches base pressure. A clean sapphire piece is mounted on the chuck and heated to 50 $^{\circ}$C above the deposition temperature. The heating step continues for an hour to degas the chamber and holder as well as remove any residual hydrocarbons from the surface of the sapphire substrate. We then decrease the temperature to the deposition temperature, which was varied from 500 $^{\circ}$C to 850 $^{\circ}$C. We adjusted the deposition rate by changing the distance between the sapphire substrate and the Ta target, as well as the dc power and chamber pressure. The chamber pressure during deposition for all these films was the same, 3 mTorr of argon. We also precondition the target before deposition by keeping the target shutter closed for a few seconds. After the Ta deposition, the Ta film is cooled down to room temperature.

Fig.~\ref{fig:PD} suggests that a slower rate in the Ta deposition tends to produce Ta films in the clean limit, while faster rates increase the defect levels and push the films to the dirty limit. For depositions at 500 $^{\circ}$C, the trend is not clear. We believe the lower temperatures lead to a worse metal-substrate interface.

\section{Transmission Electron Microscopy}
\label{TEM}
\begin{figure}
   \centering
  \includegraphics[width=0.45\textwidth]{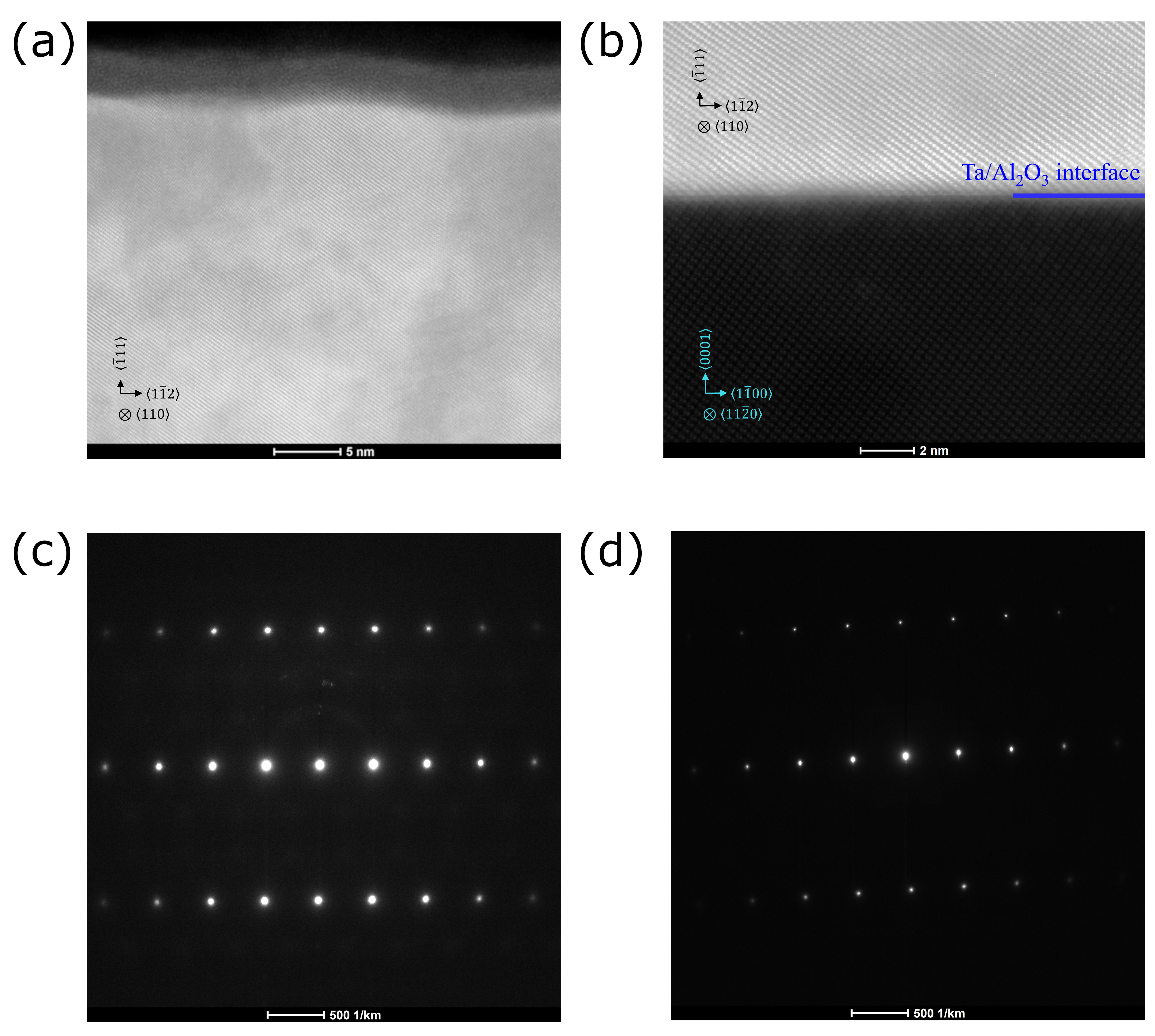}
  \caption{%
(a) Cross-sectional STEM image of a type A $\alpha$-Ta film. The estimated thickness of the amorphous oxide layer is about 3~nm. The $\alpha$-Ta has $<$111$>$ orientation. 
(b) STEM image of the $\alpha$-Ta and substrate interface. The Ta atoms are directly bonded to Al$_{2}$O$_{3}$ at the metal-substrate interface.
The diffraction patterns of Ta atoms for clean (c) and dirty (d) films are shown which confirm the cubic structure of Ta films in both types.}
   \label{fig:TEM}
\end{figure}
Scanning transmission electron microscope (STEM) images for a type A film, device D$_{1}$ (Fig.~\ref{fig:TEM}(a)) shows an amorphous oxide layer that is estimated to be about 3~nm thick, consistent with previous reports on similar Ta films~\cite{mclellan2023chemical}. The Ta film grows along the $<$111$>$ direction with a BCC $\alpha$-Ta structure consistent with our XRD results, with no evidence of the tetragonal $\beta$-Ta phase. The STEM image shows the absence of grain boundaries suggesting a single-crystal structure. Fig.~\ref{fig:TEM}(b) shows the metal-substrate interface; the Ta atoms are directly bonded to sapphire at the interface. The orientations for Ta ($<$1$\bar{1}2$$>$) and sapphire ($<$1$\bar{1}$00$>$) at the interface were used to calculate the lattice mismatch in Table~\ref{table:lattice mismatch}. Finally, the diffraction patterns in both films confirm the absence of the tetragonal ($\beta$) Ta phase (Fig.~\ref{fig:TEM}(c, d)).

\section{Scanning SQUID Microscopy}
\label{SQUID}
\begin{figure}
   \centering
  \includegraphics[width=0.5\textwidth]{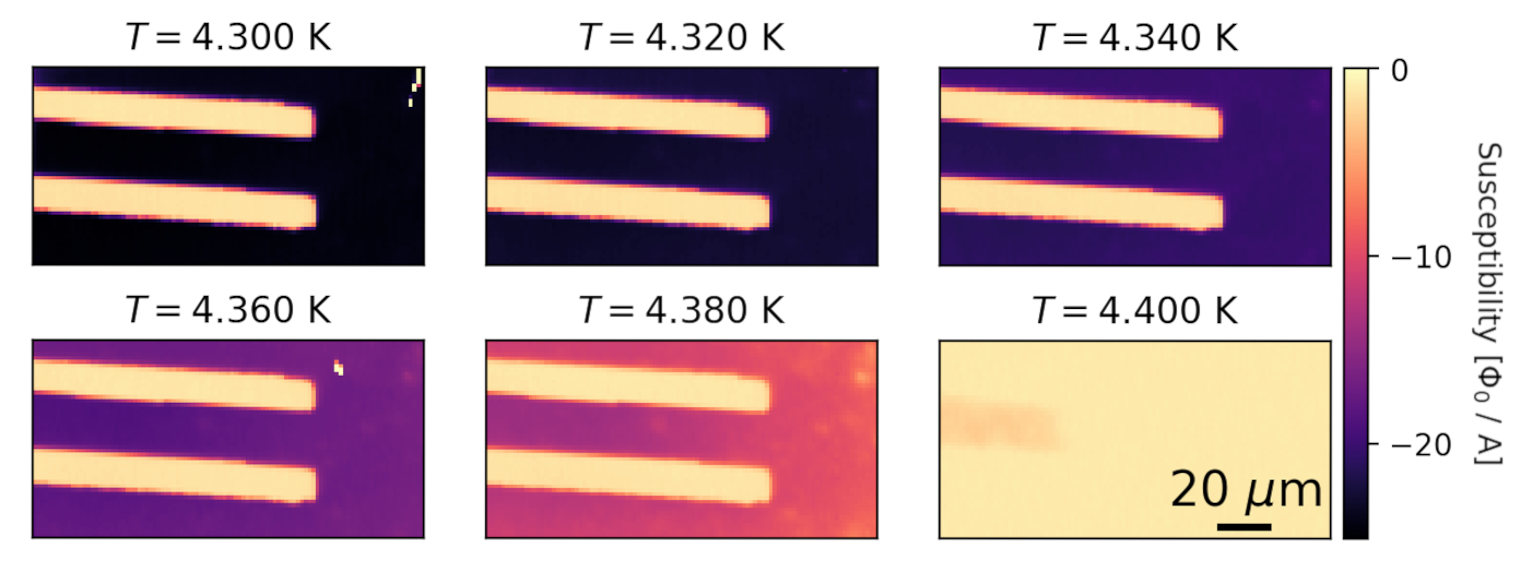}
  \caption{Temperature-dependent scanning SQUID microscopy measurements of the magnetic susceptibility. The bright areas correspond to regions of exposed sapphire substrate where Ta has been etched away. These measurements reveal a largely homogeneous magnetic response in the D$_{1}$ film, a representative type A film.
  }
   \label{fig:SQUID}
\end{figure}
Scanning SQUID microscopy can reveal inhomogeneities in Ta thin films via spatially resolved measurements of the superconducting transition temperature. Fig.~\ref{fig:SQUID} shows the magnetic susceptibility SQUID scan on a representative type A Ta film. The bright rectangular shapes are regions of exposed sapphire substrate where the Ta has been etched away. The SQUID measurement suggests a largely homogeneous critical temperature at the micrometer scale with variation of 0.01~K for this film. The T$_{c}$ measured with this technique is $\sim$4.3~K, which is consistent with the T$_{c}$ expected from $\alpha$-Ta films and rules out the presence of spatially inhomogeneous $\beta$-Ta phase domains.

\section{Upper-Limit Critical Field}
\label{Hc2}
\begin{figure}
   \centering
  \includegraphics[width=0.48\textwidth]{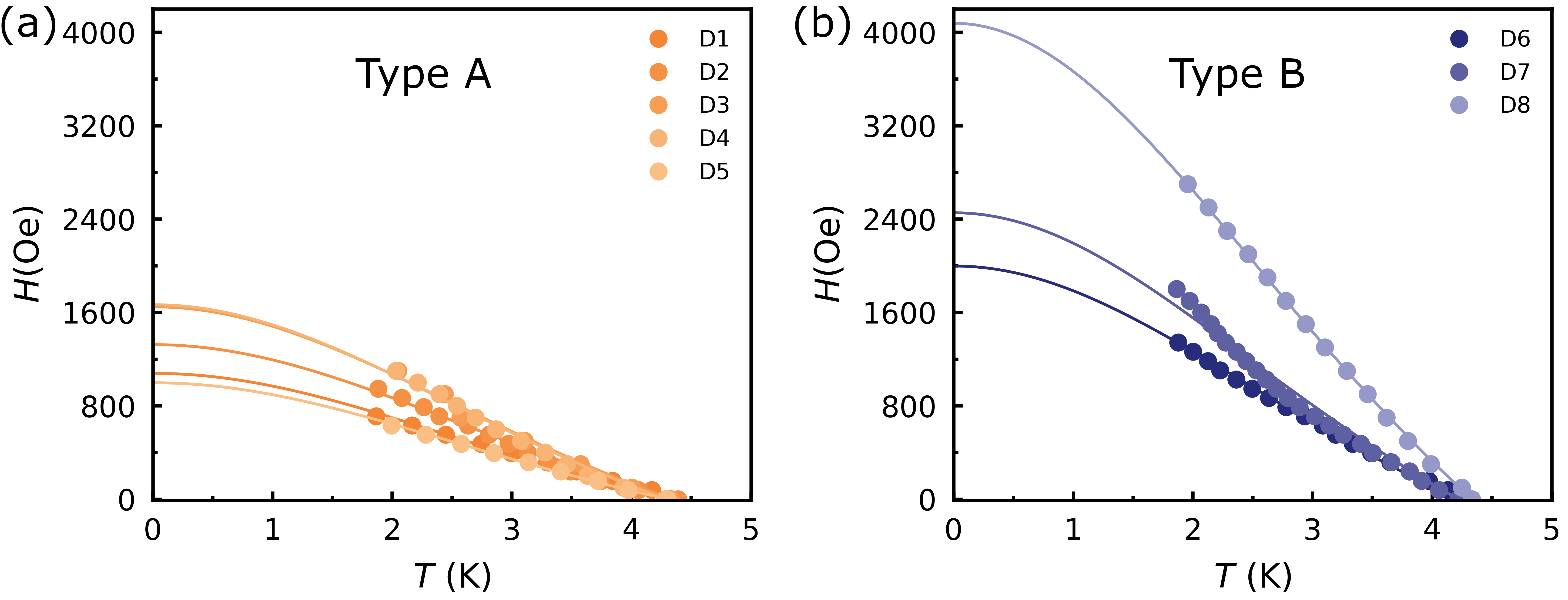}
  \caption{Applied magnetic field as  a function of extracted transition temperature phase diagram for type A (a) and type B (b) films. The solid lines represent fits obtained using Ginzburg-Landau theory to extract the upper-limit critical field for each film.
  }
   \label{fig:Hc2vsT}
\end{figure}
The magnetoresistivity as a function of temperature was measured for all the Ta devices presented in Table~\ref{table:transport_comparison}, similar to Fig.~\ref{fig:2}. We extracted T$_{c}$ as the temperature at 50\% of the superconducting transition step for each applied field strength and plotted a phase diagram for each film. The solid lines on each data set are fits based on the Ginzburg-Landau theory (Eq.~\ref{GL equ}) model; the y-intercept is the upper limit critical field (H$_{c2}$(0)). H$_{c2}$(0) is larger for type B films compared to type A films (Table~\ref{table:transport_comparison}).

\section{Mean Free Path Determination}
\label{Hall}
\begin{figure}
   \centering
  \includegraphics[width=0.4\textwidth]{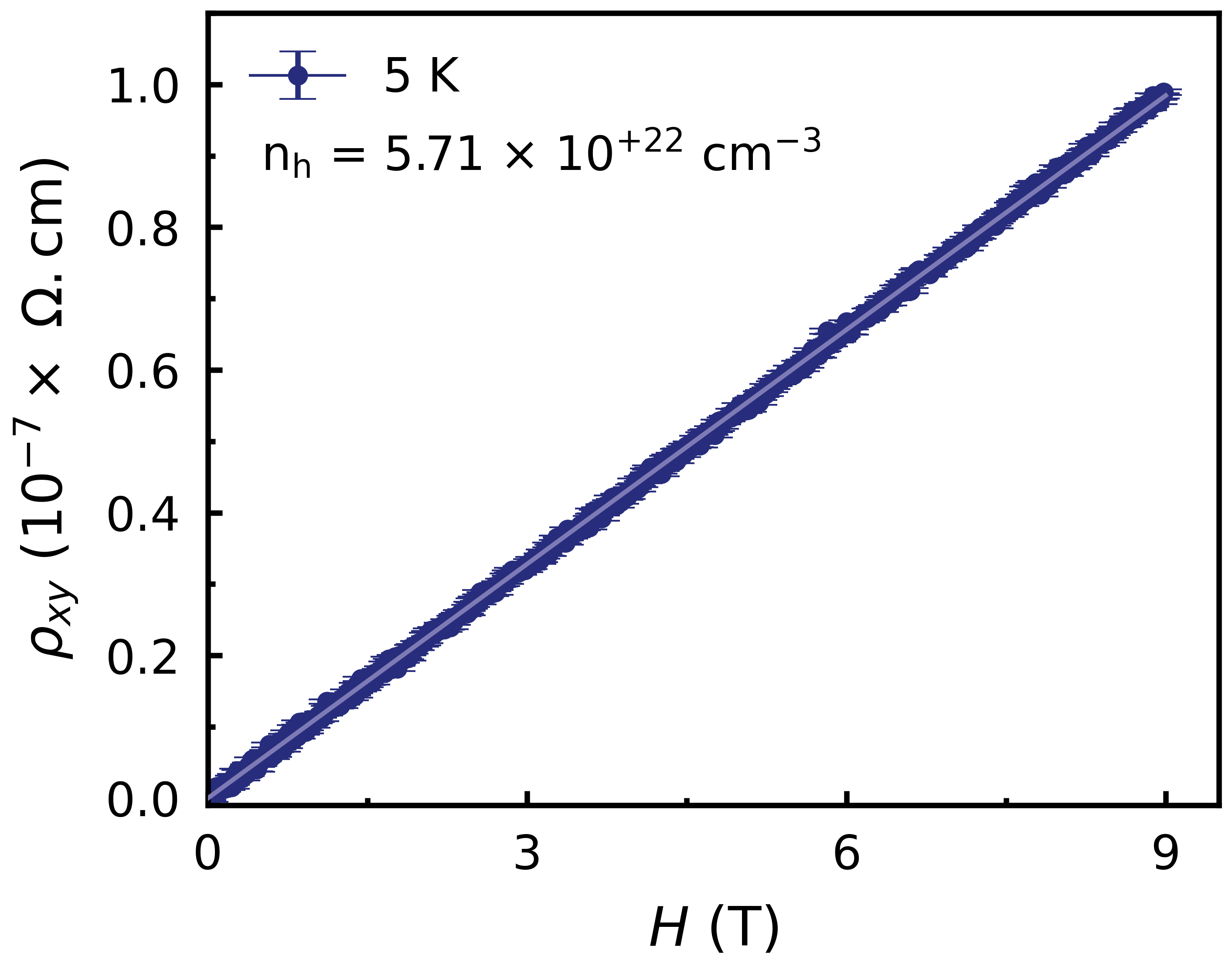}
  \caption{The Hall resistivity as a function of applied magnetic field at $T=$ 5~K for a representative Ta film. We isolated the $\rho_{xy}$ component from the $\rho_{xx}$ component by measuring the transverse resistivity in both $+\hat{z}\uparrow$ and $-\hat{z}\downarrow$ directions. The hole-charge density for the measured film is $5.71 \times 10^{+22}$ cm$^{-3}$.
  }
   \label{fig:Hall_Fit}
\end{figure}
In order to understand the superconducting limits for type A and B films we measured the transverse component of the magnetoresistivity via Hall measurements for each film. We patterned and etched a Hall bar pattern with photolithography on the Ta films and measured the Hall resistivity for each film. To isolate the transverse component ($\rho_{xy}$) from the longitudinal component ($\rho_{xx}$), we measured the Hall resistivity under perpendicular applied field in both the positive ($+\hat{z}\uparrow$) and negative ($-\hat{z}\downarrow$) directions. $\rho_{xy}$ was then calculated by anti-symmetrizing the data as:
\begin{equation}
    \rho_{xy} (|\textbf{H}|) = (\rho_{xy}^{\text{meas.}} (+|\textbf{H}|) - \rho_{xy}^{\text{meas.}} (-|\textbf{H}|))/2
\end{equation}
where $|\textbf{H}|$ and $-|\textbf{H}|$ refer to the $+\hat{z}$ and $-\hat{z}$ measurements, respectively. Fig.~\ref{fig:Hall_Fit} shows the Hall resistivity as a function of magnetic field up to 9~T. $\rho_{xy}$ is linear under increasing magnetic field with a positive slope, indicating a hole-type charge density~\cite{Matteiss1970electronic}. The slope of this linear data set provides the Hall coefficient ($R_{H} = \frac{\rho_{xy}}{H}$), and from $R_{H}$ the hole-charge density ($n_{h}$) is calculated as:
\begin{equation}
    n_{h} = \frac{1}{R_{H} e}
\end{equation}
where $e$ is the electron charge. We conducted the Hall measurements for all type A and type B films and extracted the hole-charge density for each film. The results are summarized in Table~\ref{table:hole-charge}. 
\begin{table}[h]
\centering
\caption{\label{table:hole-charge}Hole-charge density values extracted from Hall measurements for each individual film.}
\begin{ruledtabular}
\begin{tabular}{ccc}
   & Device & $n_{h}$ ($10^{+22}$ cm$^{-3}$) [5~K]   \\
       \hline
       \parbox[t]{1mm}{\multirow{5}{*}{\rotatebox[origin =c]{90}{Type A}}}
       &    D$_{1}$	&    5.893 $\pm$ 0.003  \\
       &    D$_{2}$	&    6.373 $\pm$  0.004\\
       &    D$_{3}$	&    6.976 $\pm$ 0.004  \\
       &    D$_{4}$	&    6.118 $\pm$ 0.003  \\
       &    D$_{5}$	&    6.916 $\pm$  0.005\\
       \hline\hline
       \parbox[t]{1mm}{\multirow{4}{*}{\rotatebox[origin =c]{90}{Type B}}}
       &    D$_{6}$	&    9.300 $\pm$  0.005 \\
       &    D$_{7}$	&    5.706 $\pm$  0.003 \\
       &    D$_{8}$	&    7.047 $\pm$  0.006
   \end{tabular}
   \end{ruledtabular}
\end{table}
%
The mean free path was then calculated via the relation:
\begin{equation}
    l = \frac{m_{e} v_{F}}{e^{2} n_{h} \rho_{n}}
\end{equation}
where $m_{e}$ is the electron mass, $v_{F}$ is the Fermi velocity, and $\rho_{n}$ is the normal state resistivity right above the superconducting transition temperature~\cite{paturi2022roles}. We used $v_{F} = 1.4 \times 10^{8}$ cm/s according to Ref.~\cite{warburton1995double} for Ta. The calculated mean free path values are summarized in Table~\ref{table:transport_comparison}. A film-by-film comparison between coherence length and mean free path was necessary in order to identify them as belonging to the clean or dirty limits.

\section{Surface Morphology}
\label{AFM}
\begin{figure}
   \centering
  \includegraphics[width=0.3\textwidth]{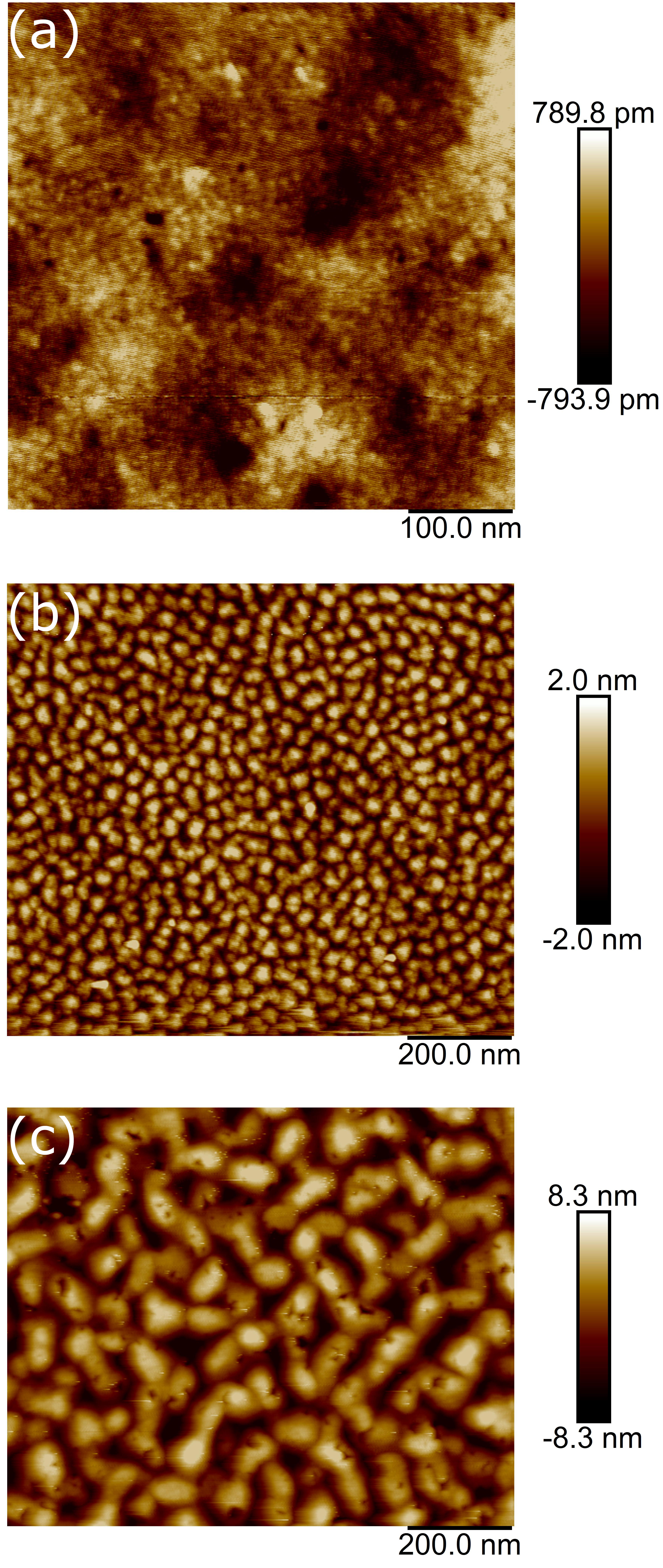}
  \caption{AFM scans of (a) type A Ta film with $<$111$>$ orientation, (b) type B film with $<$111$>$ orientation, and (c) type B film with $<$110$>$ orientation. 
  }
   \label{fig:AFM_CleanvsDirty}
\end{figure}
Surface morphology and roughness are measured by atomic force microscopy (AFM). A type A film with $<$111$>$ out-of-plane orientation shows a homogeneous surface with an average surface roughness of 0.246~nm and root mean square surface roughness of 0.198~nm (Fig.~\ref{fig:AFM_CleanvsDirty}(a)). The type B film with $<$111$>$ out-of-plane orientation has an average surface roughness of 0.652~nm and root mean square surface roughness of 0.543~nm (Fig.~\ref{fig:AFM_CleanvsDirty}(b)). We also present an example of a type B film with $<$110$>$ out-of-plane orientation. This film has an average surface roughness of 2.76~nm and root mean square surface roughness of 2.29~nm (Fig.~\ref{fig:AFM_CleanvsDirty}(c)). The increase in surface roughness and the appearance of grain-like structures for type B films compared to type A films is consistent with the higher density of defects observed for the type B films. 

\section{XRD Characterization of $<$110$>$ Oriented Film}
\label{110 orientaion}
\begin{figure}
   \centering
  \includegraphics[width=0.45\textwidth]{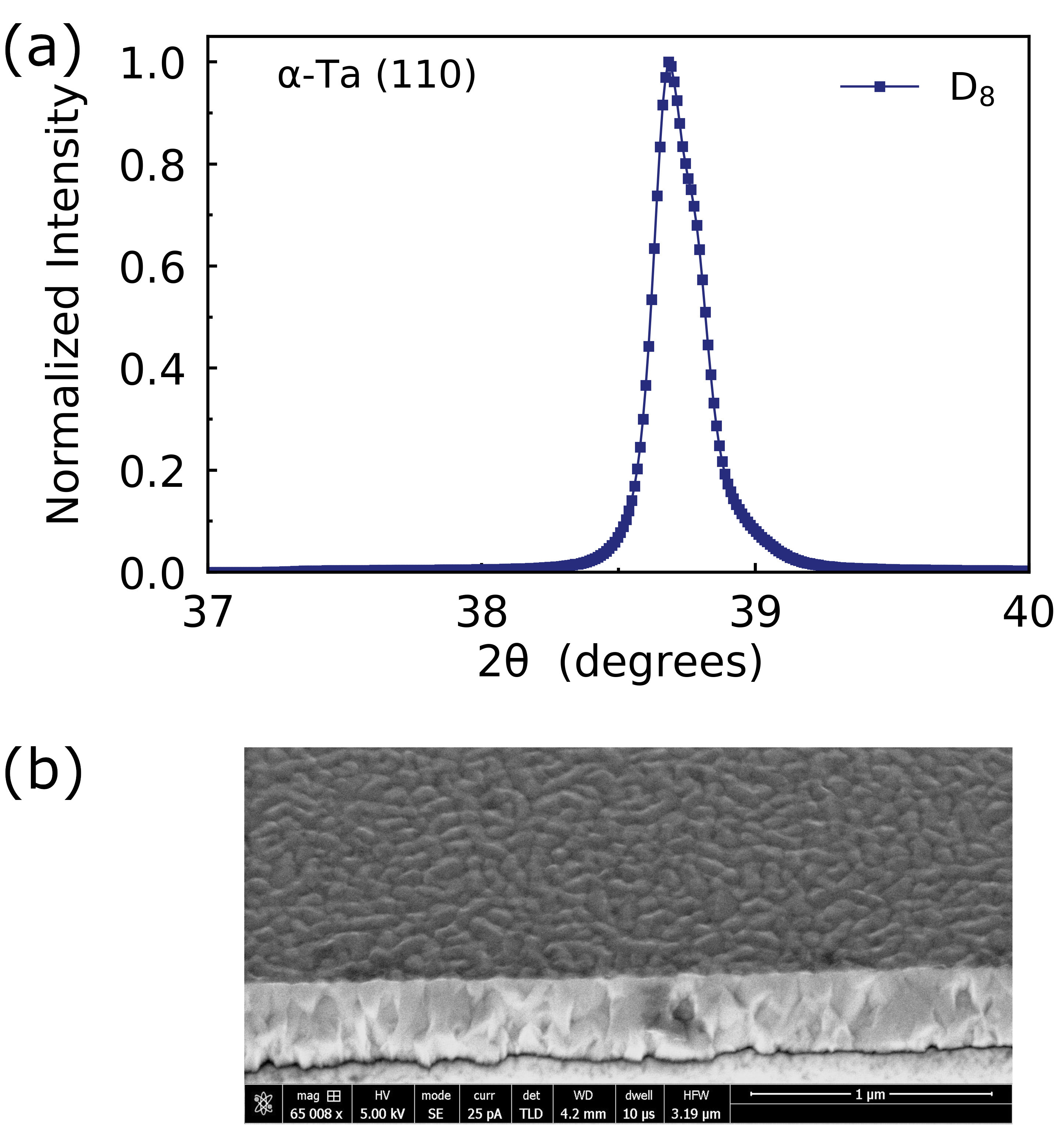}
  \caption{(a) Zoomed in XRD pattern for D$_{8}$ with $<$110$>$ out-of-plane orientation. (b) The SEM image shows the columnar structure in this $<$110$>$ $\alpha$-Ta film.
  }
   \label{fig:DEP100_SEM}
\end{figure}
We also present a representative film with an out-of-plane $<$110$>$ orientation, XRD peak at $\sim$38.6$^{\circ}$. Due to its columnar structure, the out-of-plane $<$110$>$ orientation hosts more defects compared to the $<$111$>$ orientation Ta films (Fig.~\ref{fig:DEP100_SEM}(b)). This observation is also supported by the D$_{8}$ device which has $<$110$>$ orientation and the smallest obtained RRR values among our Ta films (Table~\ref{table:transport_comparison}). The obtained lattice constant for this film is consistent with the lattice constant of the dirty limit category (Table~\ref{table:lattice mismatch}).

\section{Surface TLS Loss of Micro-structured Devices}
\label{loss}
\begin{figure}
   \centering
  \includegraphics[width=0.4\textwidth]{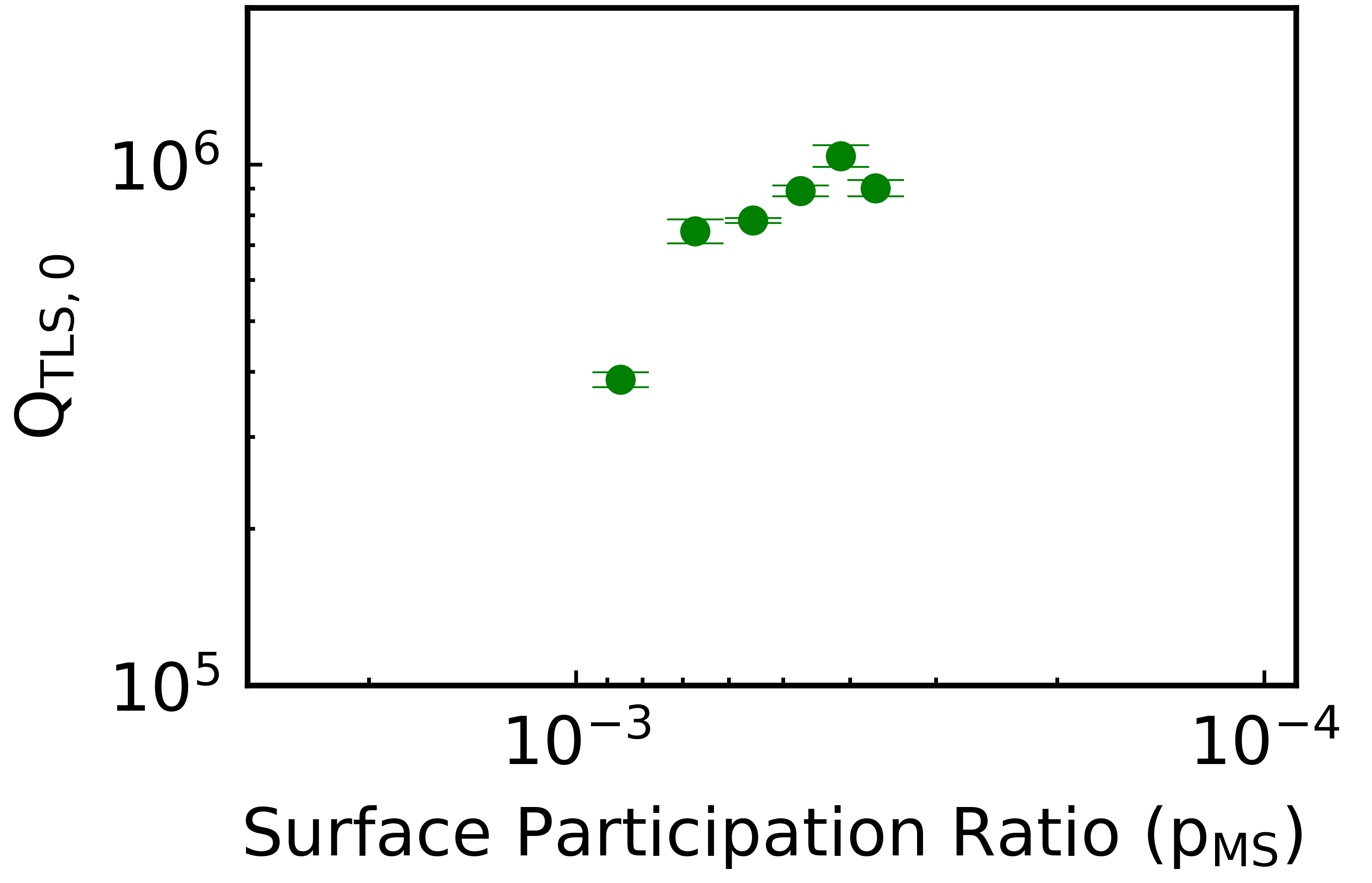}
  \caption{Dependence of the extracted Q$_{TLS, 0}$ from $\delta f/f$ measurements for the micro-structured type A device as a function of surface participation ratio. The extracted Q$_{TLS, 0}$ does not show any additional microwave loss arising from the addition of holes on the ground plane and resonator centerpin.}
   \label{wangplot}
\end{figure}

The introduction of micro-structured holes in the film suppresses vortex-motion-induced loss but can in principle increase surface TLS loss. To investigate the impact on TLS loss we measure the linear absorption due to TLSs, Q$_{TLS, 0}$ as a function of surface participation ratio for the micro-structured type A device (Fig.~\ref{wangplot}). The Q$_{TLS, 0}$ is comparable to devices of the same size without patterning \cite{crowley2023disentagling}, indicating that the additional holes do not introduce additional TLS loss.

\section{X-ray Absorption Spectroscopy}
\label{XAS}

\begin{figure}
   \centering
  \includegraphics[width=0.45\textwidth]{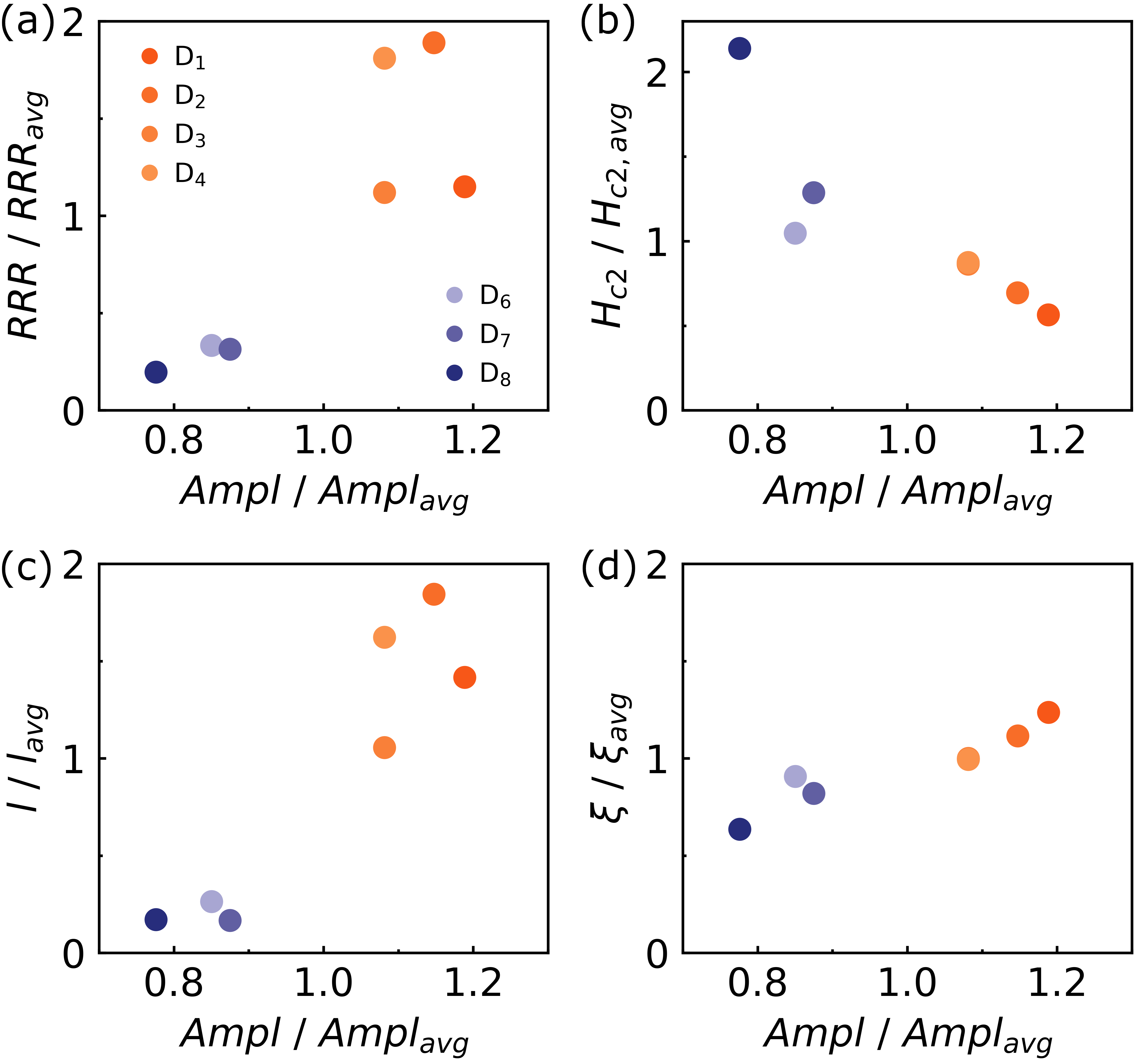}
  \caption{ A comparison between the ratio of the Ta-Ta peak amplitude from EXAFS data to (a) RRR , (b) H$_{c2}$, (c) $l$, and (d) $\xi$ for the reported devices in Table~\ref{table:transport_comparison}.
  }
   \label{fig:XAS}
\end{figure}
The local structure of both type A and B films are studied using X-ray Absorption Spectroscopy (XAS) at the 06-BMM beamline of the National Synchrotron Light Source-II (NSLS-II) at Brookhaven National Laboratory. The measurements were taken at the Ta L$_{3}$ edge using a Si (111) double crystal monochromator, with the storage ring operating at 3 GeV and a current of 400 mA. The XAS spectra were recorded at room temperature in fluorescence mode using a Vortex four-element silicon drift detector. Each sample was measured four times, and the results were merged, background subtracted and normalized using the Athena software package~\cite{ravel2005athena}. 
The XAS spectra indicate that the major component of both clean and dirty limit Ta films is metallic Ta (Ta$^{0}$). Fourier transform of the extended x-ray absorption fine structure (EXAFS) further confirmed that the films exhibit features of the pure $\alpha$-Ta phase, in agreement with previous work by Jiang et. al~\cite{jiang2003investigation}. This result was also consistent with our x-ray diffraction measurements. A distinct peak in the EXAFS data, centered at approximately 2.72~\AA in R space, corresponds to the Ta-Ta bond distance in the first coordination shell. The amplitude of this peak reflects the degree of bond ordering in the local structure~\cite{yang2012local}. 
To establish a correlation between these structural characteristics and the films’ transport properties such as: RRR, H$_{c2}$(0), $l$, and $\xi$, we plotted the ratio of the Ta-Ta peak amplitude for each sample relative to the average against the corresponding RRR, H$_{c2}$(0), $l$, and $\xi$ values (Fig.~\ref{fig:XAS}). Type A films show higher amplitude ratios compared to the type B films, a similar trend for RRR, $l$, and $\xi$ was observed.

\bibliography{bib}

\end{document}